# PHYSICAL MEANING AND EXPERIMENTAL CHECK OF A VARIATIONAL PRINCIPLE FOR MACRO-TO-MICRO TRANSITION

M. ARMINJON & D. IMBAULT
Laboratoire "Sols, Solides, Structures", Institut de Mécanique de Grenoble
B.P. 53, 38041 Grenoble cedex 9, France.

## 1. Introduction and Summary

The macroscopic behavior of a heterogeneous medium, and the microscopic fields as well (e.g. strain and stress), do not depend merely on the volume distribution of the microscopic inhomogeneity, but also on the geometrical distribution. It is often admitted that, in the ideal situation where the micro-geometrical information would be completely known, then the macroscopic behavior should be determined uniquely. According to the results of Kröner (1986), this is true for linear behavior (thus e.g. for linear elasticity).

*In this paper*, it will be argued that, for strongly non-linear behaviors such as those exhibited by inelastically deformed polycrystals, a fully deterministic position can hardly be maintained. The general reasons are the lack of information on the relevant boundary conditions, and the tendency of non-linear dynamical systems to have a "horizon of predictibility", beyond which the predictions depend too sensitively on the precise initial and boundary conditions. Thus, one may expect that *the micro-fields should possess some "stochastic" features.* This justifies a peculiarity of the variational micro-macro model proposed by Arminjon (1991a), and which may appear as a lack − namely that, in this model, the data of the microscopic behavior plus the overall stimulus has to be supplemented by a "heterogeneity parameter". In this model, the macro-to-micro transition depends on the validity of a "principle of minimal heterogeneity" (PMH). It will be proved that *the PMH has a very close relation to the maximum entropy principle* proposed by Jaynes (1957), and which is now central in statistical physics. Moreover, *two new models, based on the maximum entropy principle, will be proposed.* Lastly, the PMH will be checked by comparing a predicted *deformation texture* with the experimental one: the obtained agreement compares favourably with the agreement obtained by using the self-consistent model proposed by Molinari *et al.* (1987)

## 2. The Problem of the Boundary Conditions in Micro-Macro Models

Let us envisage a situation (a structure submitted to some boundary conditions, possibly also to volume forces) where the use of a constitutive law derived from a homogenization model makes sense, thus a situation leading to *slowly varying macroscopic fields* of strain and stress (perhaps also temperature, etc.) in the structure. It is also assumed that the structure is made of a macroscopically homogeneous material. We thus may consider that representative volume elements (RVE 's) can be defined, which, as usual, are large as compared with the typical size of the microscopic inhomogeneity, but small with respect to the structure itself, and that the stress and strain fields are macroscopically homogeneous at the scale of RVE 's: it is precisely with the relationship between the volume averages of the stress and the strain within an RVE that a homogenization model should provide us. Now in this relevant situation, the exact boundary conditions applied to any given RVE are furnished by the neighbouring ones (Hill, 1984), hence are unknown. Yet it is these conditions that should be the correct ones for the micro-macro transition! Roughly speaking, one can describe them as a kind of random fluctuation around conditions corresponding to uniform fields [more precise statements can be found in Hill (1984); cf. also Arminjon (1991b)]. We thus can state that *the exact boundary conditions that are relevant to the micro-macro transition are inherently unknown and non-unique* (since many equivalent macro-elements might be considered in a macro-homogeneous situation). Moreover, for *non-linear* behavior, one may *a priori* expect that the solution of the problem, at least the macro-to-micro transition i.e. the determination of the *microscopic fields* from the data of the overall strain or stress tensor, *may depend sensitively on the boundary conditions* that are imposed. As to the "micro-to-macro" step i.e. the determination of the overall constitutive relation by averaging the microscopic fields, an implicit postulate in any homogenization model states that, precisely: For any given value of the macroscopic stimulus **S** [some average of the microscopic one **s**(**x**), e.g. the volume average or the ensemble average], the average **R** of the response micro-field **r**(**x**) does not depend on the details of the micro-field **s**(**x**) and thus does not depend on the precise boundary conditions imposed to the field **s**(**x**). (This should be true in so far as the boundary conditions are compatible with a macro-homogeneous situation.) It seems that few rigorous results have been proved with regard to the justification of this postulate. For a *linear elastic* material, the boundary conditions corresponding to uniform *stress* or to uniform *strain* lead indeed to the same homogenized relation between the macro-stress **T** and the macro-strain **E**: this has been proved by Suquet (1982) for a periodic material and by Sab (1992) for an ergodic-random one.

In the case of non-linear behavior, however, the importance of the boundary conditions was shown in a demonstrative way for a periodic material by Turgeman and Pastor (1987). They considered a structure made of a periodic stratified composite, each stratum being a perfectly plastic Mises material. For several boundary conditions, they compared the limit loads of this structure, as calculated by two methods: in the first method, the limit load of the "real" (heterogeneous) structure was calculated. In the second method, they calculated the limit load of the structure, when filled with a homogeneous material whose behavior is that given by the periodic homogenization (Suquet, 1982) for the stratified composite. For stress-governed boundary conditions (compression or shear), the failure of the stratified structure occurred with macroscopic strain localization, which was not the case for the structure filled with the homogenized material. Moreover, *the two loads differed grossly*, even for large structures containing an increasing number of unit cells. "Macroscopic localization" means in this particular case that the velocity field was discontinuous along some slip planes traversing a *significant* part of the structure, but not the *whole* structure. (In contrast, for kinematical boundary conditions, the slip planes were rather uniformly distributed throughout the structure, and the two loads were equal.) Thus, for the stress-governed boundary conditions, the structure underwent localized strain at the macroscale, grossly inconsistent with periodic fields, and in fact with any homogenization procedure.

However, strain localization at the *microscopic* scale, e.g. a microscopic pattern of shear bands, is not incompatible with a "homogenizable" situation, provided the pattern is distributed in a macroscopically homogeneous way. In fact, one may state that, e.g. for metals, at the microscale plastic deformation is always a localized one (Krato-chvíl, 1988). The plastic strain indeed concentrates, first within crystallographically oriented slip lines, then within dense dislocation walls delineating cell blocks, still later within microscopic shear bands (Bay *et al.*, 1992; Leffers, 1993). (Eventually, also the macroscopic strain field becomes localized, from which stage one should not use a homogenization model any more.) During this process, an "ordered" pattern of intensely deformed bands is formed, though it is "random" in the details, and this pattern depends on the overall strain mode. It evolves by a continuous fragmentation into smaller microstructural units. Also for rocks, concretes and reinforced concretes, it may be said that the field of microscopic plastic strain is essentially a localized one, because in such materials the plastic strain is a combination of the plastic strain of crystalline constituents (for which similar processes as in metals are likely to operate), of slips at interfaces, and of microcracks. The same argument may be developed for composites.

This micro-localized nature of plastic deformation seems to imply that it is governed to a significant extent by microscopic *instability processes* typical of non-linear dynamical systems, as suggested for metals by Walgraef and Aifantis (1985) and by Kratochvíl (1988), and this means a significant sensitivity to the initial as well as boundary conditions[1]. Occasionally, a strong dependence of the microstructure on the boundary conditions is indeed experimentally observed (Leffers, 1993). Thus, also in "homogenizable" situations, the influence of the boundary conditions imposed in the micro-macro model cannot be neglected. In other words, one cannot just impose uniform or periodic conditions at the boundary of an RVE and admit *a priori* that the microscopic fields will be correctly representative for those in an RVE in a macrohomogeneous zone of a real deformed structure subjected to some real tractions: in view of the non-linearity, the microfields are likely to depend significantly on the imposed boundary conditions. It is even possible that the homogenized constitutive relation might itself depend on the former ones. *In summary*, the boundary conditions that would be exactly relevant to a deterministic micro-macro transition are unknown, even undetermined, and yet one cannot consider that the solution of the problem (at least the microscopic fields and the microstructure evolving as a function of these, perhaps even also the overall behavior) does not depend on the precise boundary conditions.

---

[1] A plastic deformation is an evolution problem with boundary and initial data. This may indeed be regarded as an initial value problem for a *dynamical system*, in the following way. The set of the *degrees of freedom*, $Z(t)$, should be the data of the current field of displacement, $\mathbf{u}(\mathbf{x}^0, t)$ for all $\mathbf{x}^0$ in a reference configuration $\Omega^0$ (lagrangian description), plus the current field of "state" $\mathbf{X}$ (the *state* is the set of the internal and/or geometrical variables that make the microscopic constitutive law depend on the local position). Thus, $Z(t)$ is the mapping: $\mathbf{x}^0 \to [\mathbf{u}(\mathbf{x}^0, t), \mathbf{X}(\mathbf{x}^0, t)]$ defined on $\Omega^0$. The time derivative $dZ/dt$ is hence the data of the current fields of velocity, $\mathbf{v}(\mathbf{x}^0, t)$, and of evolution rate of the state, $(d\mathbf{X}/dt)(\mathbf{x}^0, t)$. So $dZ/dt$ is determined by the current value $Z(t)$, plus the current boundary data for velocity that play the rôle of an *external forcing*, and one indeed has a "dynamical system" $dZ/dt = f(Z(t), t)$, though it is not given in explicit form. In the modern theory of non-linear dynamical systems and chaos, a *dynamical*

## 3. Inhomogeneous Variational Model and Principle of Minimal Heterogeneity

In the inhomogeneous variational model (IVM), the micro-macro transition depends on the volume distribution of the states (of the different constituents) *and* on the average inhomogenity $r_0$ of the *actual* local stimulus field (e.g. the average difference between the overall strain and the strains in the different constituents). Clearly, this parameter depends on the pattern of microscopic strain localization and represents an information which is consistent with the foregoing analysis of the reasons that make the need for additional information. Let us recall the key points in the IVM (Arminjon, 1991a):

Instead of the true micro-fields [e.g. strain-rate **d**(**x**) and stress **t**(**x**)] depending on the spatial position **x**, "constant-state averaged" local fields are considered (the local state **X** determines the local behavior, see note [1]). If only a finite number of different states, $\mathbf{X}_1, ..., \mathbf{X}_n$ are present (which may always be assumed as an approximation as good as one wishes), the *constituents* $\Omega_1, ..., \Omega_n$ may be defined as the *zones with constant* **X**, with volume fractions $f_k$ ($f_1 + ...+ f_n = 1$). Note that a such zone $\Omega_k$ will in general involve many separated domains, e.g. grains with the same orientation. The unknown of the macro-to-micro transition is then the distribution $(\mathbf{D}_k)_{k = 1, ..., n}$ of the average strain-rates in the constituents, for a given value of the macroscopic strain-rate **D**. The IVM applies to any behavior which derives from a convex potential $u$, thus $u_1$, ..., $u_n$ in $\Omega_1, ...,\Omega_n$. It consists in searching the following minimum of the average potential with two constraints, depending on the given **D** and on the heterogeneity parameter $r$:

---

*system* is nothing else than a *differential system*, independently of whether acceleration terms are neglected or not. Thus, even a *quasi-static* deformation process leads, as we have just explained, to a dynamical system – a non-linear one for a non-linear constitutive equation.

$$U_r(\mathbf{D}) \equiv \mathrm{Min}\,[f_1\, u_1(\mathbf{D}^*_1) + \ldots + f_n\, u_n(\mathbf{D}^*_n)] \text{ under constraints } \mathbf{D}^* = \mathbf{D} \text{ and } h \leq r. \quad (1)$$

Here $\mathbf{D}^* \equiv f_1\, \mathbf{D}^*_1 + \ldots + f_n\, \mathbf{D}^*_n$ is the macro-average of the distribution $(\mathbf{D}^*_k)_{k=1,\ldots,n}$, and

$$h = h((\mathbf{D}^*_k)) \equiv [f_1\, \|\mathbf{D}^*_1 - \mathbf{D}^*\|^p + \ldots + f_n\, \|\mathbf{D}^*_n - \mathbf{D}^*\|^p]^{1/p} \quad (2)$$

is the definition of the average heterogeneity [with $p = 1$ for rate-independent plasticity, Arminjon *et al.* (1995)]. Depending on $r$, the minimum problem (1) makes a continuous transition from Voigt's model ($r = 0$) to Reuss' model [$r \geq R$ for some value $R = R(\mathbf{D})$]; for any $\mathbf{D}$, there exists a generically unique value $r_0(\mathbf{D})$ such that $U_{r_0}(\mathbf{D}) = U(\mathbf{D})$, the exact value of the macro-potential (Arminjon, 1991a). For the *micro-to-macro* transition, the mere problem is to find $r_0$. One postulates a simple dependence $r_0 = r_0(\mathbf{D})$, e.g. $r_0 = a\, \|\mathbf{D}\|$, and one "adjusts" $a$ from *one* mechanical test (Arminjon *et al.*, 1994).

For the *macro-to-micro transition*, one has to assume that the distribution solution of the minimum problem (1), $(\mathbf{D}^{\mathrm{sol}}_k)$ [for the relevant value, $r_0 = r_0(\mathbf{D})$, of the heterogeneity parameter], is the actual distribution $(\mathbf{D}_k)$. This is equivalent to assuming that $r_0$ is precisely the heterogeneity of the actual distribution $(\mathbf{D}_k)$, and in turn it amounts (Arminjon *et al.*, 1994) to assuming the following *principle of minimal heterogeneity* (PMH) for the actual distribution:

Among distributions $(\mathbf{D}^*_k)$ that have the relevant macro-average, $\mathbf{D}^* = \mathbf{D}$, and that lead to the correct value $U(\mathbf{D})$ of the average potential, i.e., that are such that

$$\langle u \rangle \equiv f_1\, u_1(\mathbf{D}^*_1) + \ldots + f_n\, u_n(\mathbf{D}^*_n) = f_1\, u_1(\mathbf{D}_1) + \ldots + f_n\, u_n(\mathbf{D}_n) \equiv U(\mathbf{D}), \quad (3)$$

the actual distribution $(\mathbf{D}_k)$ has the least heterogeneity $h$.

The PMH allows successful predictions of deformation textures for steels (Arminjon & Imbault, 1996), but it cannot be derived from deterministic mechanics. Instead, we shall establish below a link between the PMH and the maximum entropy principle.

## 4. The Principle of Maximum Statistical Entropy (MAXENT) in Statistical Physics

The MAXENT principle (Jaynes, 1957) gives the link between information theory and statistical mechanics. Information theory leads unambiguously to the following expression, called *statistical entropy*, for the "amount of uncertainty" represented by a probability distribution $(p_i)_{i=1,\ldots,M}$ on a finite set $\mathrm{E} = \{x_1, \ldots, x_M\}$:

$$S = -K \sum_i p_i\, \mathrm{Log}\, p_i, \quad (4)$$

with $K$ an arbitrary positive constant. Consider the situation where only some expectation values

$$\langle \phi_q \rangle \equiv \sum_i p_i\, \phi_q(x_i) = a_q \qquad (q = 1, \ldots, Q), \quad (5)$$

are known (with $\phi_q$ *known* functions, $Q < M$, in practice $Q \ll M$). This does not determine the distribution ($p_i$). The MAXENT principle selects the distribution ($p_i$) that maximizes $S$ with the $Q$ constraints (5). It amounts to *selecting the **broadest** probability distribution compatible with the available information* (Jaynes 1957). I.e., the "unbiased choice". As shown by Jaynes (1957), this principle is the *only* necessary tool to derive *all* the laws of statistical physics such as Boltzmann's distribution, etc. [see Balian (1991) for a detailed proof of the latter assertion]. According to this view, the "physics" in statistical physics may be reduced to the mere enumeration of the different possible micro-states.

But, even before one may enumerate the different micro-states, the application of the MAXENT procedure needs to set the relevant physical problem in some probabilistic framework. In statistical physics, the probabilities appear because statistical ensembles of "macroscopically identical, but microscopically different systems" are considered: $p_i$ is the probability that a "randomly selected" *system* is in the global micro-state (*i*). A *global* micro-state is the data of the micro-states of all elementary constituents (e.g. particles) – an *elementary* micro-state being e.g. the set of the quantum numbers of a particle (in quantum statistical mechanics) or its position and velocity (in classical SM). A macro-state is then a very formal object: it is the data of a probability distribution ($p_i$) on the space E of the global micro-states (Balian, 1991) (we assume here that E is discrete, for simplicity). However, this distribution, i.e. the macro-state, is defined only in an incomplete way, by the macroscopic (statistical) constraints (5), which correspond to assigning given average values to macroscopic parameters such as pressure and density (the average being over a statistical ensemble). In this context, the macro-state is *a priori* an undefined concept: the MAXENT principle is merely the *choice* of a particular distribution ($p_i$) (the unbiased one i.e. the broadest) in order that the macro-state be defined from the data of the macroscopic parameters. The impressing success of this par-ticular choice, and indeed of statistical mechanics, is then explained by the fact that the relevant huge numbers lead, "for the macroscopic quantities actually measured, to enor-mously sharp peaks", so that the particular distribution does not really matter, all other reasonable choices giving very nearly the same macroscopic predictions (Jaynes, 1957).

**5. A Formulation of the MAXENT Principle for Heterogeneous Continua**

We shall adopt a different framework (the "realism"), according to which the global micro-state of any *given* system is *well-defined* at any given time (and, of course, evolves with time). For us, the index *i* refers to an *elementary* micro-state, and $p_i$ (resp. $l_i$) is simply the fraction (resp. the number) of the elementary constituents that are currently in that micro-state (*i*), thus

$$p_i = l_i/N \qquad (6)$$

if $N$ is the (very large) number of elementary constituents. In mechanics and physics of heterogeneous media, a *system* will be an RVE in a statistically homogeneous medium, e.g. a polycrystal. The *constituents* $\Omega_k$ ($k = 1, ..., n$) were defined in Sect. 3. Each of them is subdivided into a large number $N_k$ of *elementary constituents*. All of these have the same volume, independent of the constituent ($k$). The volume fractions $f_k \equiv N_k/N$ are known. The *micro-state* of an elementary constituent will be defined as the joined data (**s**, **X**) with **s** the volume average of the "stimulus" (i.e., primary) micro-field in the *elementary* constituent and **X** the *state* of the corresponding constituent (Note 1). The geometry of the subdivision in elementary constituents is quite arbitrary, but it plays no rôle. To fix the ideas, **X** will be the crystal orientation **R** and **s** will be the strain-rate **d**, in a plastically deformed polycrystal. We assume that, for a given value of the macroscopic stimulus **D**, **d** may take on $m$ different values $\mathbf{D}^j$, centered at **D**. Thus

$$\mathbf{R} \in \{\mathbf{R}_1, ..., \mathbf{R}_n\}, \quad \mathbf{d} \in \{\mathbf{D}^1, ..., \mathbf{D}^m\}, \quad (\mathbf{D}^1 + ... + \mathbf{D}^m)/m = \mathbf{D}. \tag{7}$$

This assumption, if interpreted physically, would mean that the strain-rate should take "quantized" values! It is merely a convenient discretization, which may be refined at will (by increasing $m$). Note that the index $i$ of the micro-state becomes a double one, ($j$, $k$). Let $l_i = l^j{}_k$ the number of elementary crystals with micro-state ($\mathbf{D}^j, \mathbf{R}_k$). We have the constraint:

$$\sum_{j=1}^{m} l^j{}_k = N_k \quad \text{or} \quad \sum_{j=1}^{m} p^j{}_k = f_k \quad (k = 1, ..., n), \quad p^j{}_k \equiv l^j{}_k/N, \tag{8}$$

because any elementary crystal of a given orientation $\mathbf{R}_k$ must have some strain-rate $\mathbf{D}^j$. This expresses the fact that the volume fractions of the orientations, $f_k$, are known, and it implies the normalization of the discrete probability law ($p^j{}_k$). The average strain-rate in the orientation $\mathbf{R}_k$ is

$$\mathbf{D}_k = (l^1{}_k \mathbf{D}^1 + ... + l^m{}_k \mathbf{D}^m)/N_k = (p^1{}_k \mathbf{D}^1 + ... + p^m{}_k \mathbf{D}^m)/f_k. \tag{9}$$

Hence, the "**D*** = **D**" condition (that the macroscopic average of the strain-rate is the given **D**, Sect. 3) is written as

$$\sum_{j,k} p^j{}_k \mathbf{D}^j = \mathbf{D}. \tag{10}$$

The last constraint to be imposed on the probability distribution ($p^j{}_k$) is that also the average potential per unit volume is assigned to have the actual macroscopic value:

$$\langle u \rangle \equiv \sum_k f_k\, u_k(\mathbf{D}_k) \equiv \sum_k f_k\, u_k\left(\sum_j p^j{}_k \mathbf{D}^j / f_k\right) = U(\mathbf{D}). \tag{11}$$

This constraint may be surprising at first: when the macroscopic stimulus (here **D**) is known, the macro-potential should be known. But this assumes that the macroscopic constitutive law is determined, and it cannot be true as long as we only know the volume fractions of the constituents. Thus, Eq. (11), i.e. the data of the average potential, is the minimum information we must add in order to determine the macroscopic behavior − and this is now indeed determined in favourable ("statistically homogeneous") situations, because then the average potential is indeed a potential for the macro-law (see Arminjon (1991a) and references therein). However, the MAXENT principle *may* be applied with Eqs. (8)$_2$ and (10) as the only constraints: although this leads to the somewhat unphysical conclusion that the macro-law depends merely on the volume fractions of the constituents, it may be a useful approximation for weakly heterogeneous media. This simple model [MAXENT with constraints (8)$_2$ and (10)] describes a "maximally disordered" medium, in the sense that it will lead to the highest possible value for *S*. This reminds of self-consistent models, which, in the case of linear behavior, are likely to describe "perfectly disordered" media, the notion of perfect disorder referring to the geometrical distribution of the material inhomogeneity (Kröner, 1978). In the new "simple model", however, the measure of disorder is the statistical entropy of the strain-rate distribution (more generally, the statistical entropy of the distribution of the "stimulus" field among elementary constituents).

In summary, the MAXENT principle determines the distribution ($p^j{}_k$) as the one that maximizes the statistical entropy $S = - \Sigma_{j,k} \ p^j{}_k \ \text{Log} \ p^j{}_k$ under constraints (8)$_2$, (10) *and* (11) [*"general model"*, that may describe any "degree of order"], or under constraints (8)$_2$ and (10) [*"simple model"*, describing a "maximally disordered" material]. Once we know ($p^j{}_k$), we may calculate the average strain-rates per orientation, $\mathbf{D}_k$, Eq. (9). From these, we may calculate the current average rotation rates and update the orientations after a small deformation step, i.e., we may calculate the texture evolution in the same way as in Arminjon & Imbault (1996). It should be clear, however, that the two new models may be used also in much more general situations. In particular, the "simple model" does not even depend on the existence of a potential for the constitutive law.

## 6. Minimum Heterogeneity versus Maximum Statistical Entropy

Considering again, to fix the ideas and in preparation for the experimental comparison, a plastically deformed polycrystal, let us compare two models that determine the distribu-tion of the average strain-rates $\mathbf{D}_k$ in the *constituents* (each of which is the non-contiguous zone made of all grains with a given crystal orientation): (i) the "general model" maximizes the statistical entropy $S = - \Sigma_{j,k} \ p^j{}_k \ \text{Log} \ p^j{}_k$ with constraints (8)$_2$, (10) and (11), and then obtains the $\mathbf{D}_k$ 's by Eq. (9); (ii) the inhomogeneous variational model (IVM) obtains directly the $\mathbf{D}_k$ 's by solving the minimum problem (1). As we recalled at the end of Sect. 3, the IVM is *strictly* equivalent to minimizing the heterogeneity (2) of the strain-rate distribution under constraints "$\mathbf{D}^* = \mathbf{D}$" and (3), i.e. "$\langle u \rangle = U$ ". The "general model" is based on distributions ($\mathbf{D}_k^*$) satisfying both constraints "$\mathbf{D}^* = \mathbf{D}$" and "$\langle u \rangle = U$", since these constraints, expressed in terms of the $p^j{}_k$'s, are nothing else than the constraints (10) and (11), respectively.

Thus, the "general model" maximizes the statistical entropy of the $(p^j{}_k)$ distribution under the constraints "$\mathbf{D}^* = \mathbf{D}$", "$\langle u \rangle = U$" *and* the constraint $(8)_2$, that the volume fractions of the different orientations are given (a constraint that is automatically taken into account in the IVM). In other words, the constraints of both models are the same, except for the fact that the unknown $(p^j{}_k)$ of the "general model" is at a lower level than the unknown $(\mathbf{D}_k)$ of the IVM. Now, maximizing the statistical entropy with certain constraints amounts to selecting the broadest probability distribution compatible with the constraints, i.e. the distribution *closest to the uniform distribution* (Jaynes, 1957). In the present case, the distribution $(p^j{}_k)$ must in particular satisfy $(8)_2$, i.e. $\Sigma_j \, p^j{}_k = f_k$ for all $k$. Hence, the MAXENT principle selects, in average over $k$ and accounting for the remaining constraints, i.e., "$\mathbf{D}^* = \mathbf{D}$" and "$\langle u \rangle = U$", the $(p^j{}_k)$ distribution closest possible to the distribution that is uniform at fixed $k$, i.e., $p^j{}_k = f_k/m$. But, reexpressing the heterogeneity $h$ [Eq. (2)] with the help of Eq. (9), we get:

$$h^p \equiv \sum_{k=1}^{n} f_k \left\| \mathbf{D}_k - \mathbf{D} \right\|^p = \sum_{k=1}^{n} \left\| \sum_{j=1}^{m} p^j{}_k \mathbf{D}^j - f_k \mathbf{D} \right\|^p / f_k{}^{p-1}, \qquad (12)$$

and for $p^j{}_k = f_k/m$, this is simply zero! Thus, the $(p^j{}_k)$ distribution closest possible to the distribution $p^j{}_k = f_k/m$ is the distribution that leads to the minimum value of the heterogeneity $h$ (of course, minimizing $h$ is equivalent to minimizing $h^p$ with any fixed $p \geq 1$). We have thereby proved that the "general model" based on the MAXENT principle is equivalent, from a *physical* (hence approximate) point of view, to the inhomogeneous variational model. In our opinion, this reinforces the physical basis of the latter. Whether the two models are *exactly* (mathematically) equivalent, i.e. whether both predict *exactly the same* distribution $(\mathbf{D}_k)$, is an open question. However, the exact equivalence could be true only for a specific value of the number $p$.

**7. Experimental Check of the IVM: the Cold-Rolling Texture of Low-Carbon Steel**

Fig. 1. Standard skeleton lines of the 75% cold-rolling texture of a low carbon steel, as measured (exp), or calculated either with the VPSC model or with the in-homogeneous variational model ($a = 0.25$: ivm r = 0.25). *X-axes:* angle in degrees. *Y-axes:* orientation density. Truncation order $l_{max} = 22$, Gauss angle $\phi_0 = 8°$. We are grateful to F. Wagner and J.J. Fundenberger (LETAM, Université de Metz) who made the measurement and the VPSC calculation.

The prediction of the crystallographic texture of low-C steels after cold-rolling is a severe test for polycrystal models, as it has been discussed in some detail in previous works (see Arminjon & Imbault, 1996, and references therein). On *Fig. 1* here, we compare for the first time the prediction of the viscoplastic self-consistent (VPSC) model, proposed by Molinari *et al.* (1987), to that of the IVM. The general figure of the texture being very well predicted by both models, the precise comparison is based on the "skeleton lines", the (standard) definition of which may be found e.g. in Arminjon & Imbault (1996), and that contain the main preferred orientations. Both models overestimate the sharpness of the texture. The IVM predicts more efficiently the position of the peak orientations, and the ratios between the values of the orientation density on these peak orientations, than does the VPSC model. This will be detailed in future work.